\documentclass[3p,times]{elsarticle}

\usepackage{ecrc}

\usepackage{ulem}

\volume{00}

\firstpage{1}

\runauth{}

\usepackage{amssymb}
\usepackage{multirow}
\usepackage[figuresright]{rotating}

\begin{document}

\begin{frontmatter}

\dochead{}
\title{Experimental search for the violation of Pauli Exclusion Principle}

\author[label1,label2]{H.~Shi\corref{cor1}\fnref{fn1}}
\ead{Hexi.Shi@oeaw.ac.at} 
\author[label3]{E.~Milotti\corref{cor1}}
\ead{Edoardo.Milotti@ts.infn.it} 
\author[label1]{S.~Bartalucci}
\author[label1]{M.~Bazzi}
\author[label4]{S.~Bertolucci}
\author[label5,label1]{A.M.~Bragadireanu}
\author[label2,label1]{M.~Cargnelli}
\author[label1]{A.~Clozza}
\author[label1]{L.~De Paolis}
\author[label6]{S.~Di Matteo}
\author[label7]{J.-P.~Egger}
\author[labbel8]{H.~Elnaggar}
\author[label1]{C.~Guaraldo}
\author[label1]{M.~Iliescu}
\author[label9]{M.~Laubenstein}
\author[label2,label1]{J.~Marton}
\author[label1]{M.~Miliucci}
\author[label2,label1]{A.~Pichler}
\author[label5,label1]{D.~Pietreanu}
\author[label10,label1]{K.~Piscicchia}
\author[label1]{A.~Scordo}
\author[label1,label5]{D.L.~Sirghi}
\author[label1,label5]{F.~Sirghi}
\author[label1]{L.~Sperandio}
\author[label11,label1]{O.~Vazquez Doce}
\author[label2]{E.~Widmann}
\author[label2,label1]{J.~Zmeskal}
\author[label1,label5,label10]{C.~Curceanu}

\address[label1]{INFN, Laboratori Nazionali di Frascati, C.P. 13, Via E. Fermi 40, I-00044 Frascati(Roma), Italy,}
\address[label2]{Stefan-Meyer-Institut f\"{u}r Subatomare Physik, Boltzmanngasse 3, 1090 Wien, Austria,}
\address[label3]{Dipartimento di Fisica, Universit\`{a} di Trieste and INFN-Sezione di Trieste, Via Valerio, 2, I-34127 Trieste, Italy,}
\address[label4]{Dipartimento di Fisica e Astronomia, Universit\`{a} di Bologna, Viale Berti Pichat 6/2, Bologna, Italy,}
\address[label5]{IFIN-HH, Institutul National pentru Fizica si Inginerie Nucleara Horia Hulubbei, Reactorului 30, Magurele, Romania,}
\address[label6]{Universit\'{e} de Rennes, CNRS, IPR (Institute de Physique de Rennes), UMR 6251, F-35000 Rennes, France,}
\address[label7]{Institut de Physique, Universit\'{e} de Neuch\^{a}tel, 1 rue A.-L. Breguet, CH-2000 Neuch\^{a}tel, Switzerland,}
\address[label8]{Debye Institute for Nanomaterial Science - Utrecht University, P.O. Box 80.000 3508 TA Utrecht, The Netherlands,}
\address[label9]{INFN, Laboratori Nazionali del Gran Sasso, S.S. 17/bis, I-67010 Assergi (AQ), Italy,}
\address[label10]{Museo Storico della Fisica e Centro Studi e Ricerche Enrico Fermi, Piazza del Viminale 1, 00183 Roma, Italy,}
\address[label11]{Excellence Cluster Universe, Technische Universit\"{a}t M\"{u}nchen, Boltzmannstra\ss e 2, D-85748 Garching, Germany.}

\cortext[cor1]{Corresponding authors,}
\fntext[fn1]{Current address : Institut f\"{u}r Hochenergiephysik der \"{O}sterreichischen Akademie der Wissenschaften, Nikolsdorfer Gasse 18, A-1050 Wien, Austria.}


\begin{abstract}
The VIolation of Pauli exclusion principle -2 experiment, or VIP-2 experiment, at the Laboratori Nazionali del Gran Sasso 
searches for x-rays from copper atomic transitions that are prohibited by the Pauli Exclusion Principle. 
Candidate direct violation events come from the transition of a $2p$ electron to the ground state that is already occupied by two electrons. 
From the first data taking campaign in 2016 of VIP-2 experiment, 
we determined a best upper limit of 3.4$~\times~$10$^{-29}$ for the probability that such a violation exists. 
Significant improvement in the control of the experimental systematics was also achieved, although not explicitly reflected in the improved upper limit. 
By introducing a simultaneous spectral fit of the signal and background data in the analysis, 
we succeeded in taking into account systematic errors that could not be evaluated previously in this type of measurements. 
}
\end{abstract}
 

\end{frontmatter}

\graphicspath{{Figures/}}

\section{Introduction}
The Pauli Exclusion Principle (PEP) explains in a beautiful and elegant way the stability of atoms and a host of other phenomena, 
but when it was first stated by Pauli in 1924 it was met with skepticism by Bohr and Heisenberg, and  Pauli himself thought that his own construction -- 
which included electron spin as well as the exclusion principle -- could not easily find support in common sense \cite{Cha02}. Nowadays the spin-statistics connection -- on which PEP is based -- 
is one of the basic results of Quantum Field Theory (QFT), and PEP stands as one of the great cornerstones of physics. The principle is extremely robust, 
and its validity within QFT is unchallenged (see, e.g., \cite{Gre89}).

\medskip

However, QFT is not complete \cite{Jac99} and current proofs of the spin-statistics connection may possibly break in extensions of QFT, 
paving the way to the need of experimental verification of such important results of the theory like the spin-statistics connection and PEP.

\medskip

In principle, violations of PEP can be large, e.g., where there are macroscopic violations of statistics of bosons and fermions, or small, if particles violate PEP ``only a little''. 
Exhaustive reviews of the experimental and theoretical searches for a small violation of  PEP can be found in \cite{Bel10} and \cite{Ell12}.

\medskip

Particle identity -- and therefore the necessity of Hamiltonians that are symmetric with respect to particle exchange -- rules out any violation of PEP in closed systems, 
where the Messiah-Greenberg superselection rule \cite{Mes64} holds (no transitions allowed between states with different symmetry). 
This means that transitions between different symmetry states can only be searched in open systems, 
and the prototype experiment of this class has been first carried out several years ago by Ramberg and Snow \cite{Ram90}. 
In the experiment they injected a high electric DC current in a copper conductor, 
and they searched for x-rays from transitions that are PEP-forbidden after electrons are captured by copper atoms. 
In particular, they searched for PEP-violating transitions from the $2p$ level to the $1s$ level of the copper atoms, which is already occupied by two electrons. 
Because of the shielding effect of the additional electron in the ground level, the energy of such abnormal transitions deviates from the copper $K_{\alpha}$ x-ray at 8 keV by about 300 eV, 
as shown in the Appendix.
They can be distinguished in precision spectroscopic measurements using silicon detectors with typical full width half maximum (FWHM) energy resolution less than 200 eV at 8 keV. 
If we assume as did Ramberg and Snow, that the ``new'' electrons injected by the external current source have no prior established symmetry with the electrons inside the copper atoms, 
the detection of the energy-shifted x-rays is an explicit indication of the violation of spin-statistics, and thus of the violation of the PEP for electrons.

The probability of a small violation is represented conventionally in the form of a small violation probability $\beta^2/2$ parameter. 
First introduced by Ignatiev and Kuzmin \cite{Ign87} in a model to accommodate a fermionic state with multi-particle occupancy, 
the $\beta$ parameter stayed, although the model turned out to be impossible to extend into a full relativistic QFT, 
and a self-consistent theoretical framework is still missing, as reviewed by Okun \cite{Oku89} and by Greenberg \cite{Gre00}. 
In the absence of a complete relativistic theory, the only alternative view compatible with Quantum Mechanics is that there exist rare states of wrong symmetry, 
as discussed by Rahal and Campa \cite{Rah88}. 
If these states exist, $\beta^2/2$ can be interpreted as the probability of having the wrong symmetry fermions meet. 
When this happens, and the fermion system is not in its ground state, we should be able to observe fast electromagnetic transitions with the emission of light quanta. 
In this framework (as discussed by Okun in his 1989 review paper \cite{Oku89}) there can be only two kinds of experiments: searches for non-Paulian atoms and nuclei, 
or searches for anomalous electromagnetic transitions. 

Interpretations of the results from different types of measurement can be model-dependent. 
Without a self-consistent theory, even phenomenologically, to describe a small violation of PEP, 
a comparison of the limits of the violation probability out of a thorough review of the experimental context can be misleading and controversial. 
For example, the most stringent upper limit of the $\frac12\beta^2$ is set by Borexino \cite{Bel10} at the order of 10$^{-60}$, 
derived from a null result of measuring the transition of a stable $^{12}$C nuclei into an abnormal $^{11}\tilde{B}$ nuclei with 3 protons or neutrons in the S shell, meanwhile releasing a proton. 
For this type of measurement, one can argue that it is the stability of matter that is under test, not the PEP itself. 
Since in a stable system, the symmetrization group it belongs to, either normal fermionic or possibly mixed, 
has been established and will not change due to the more fundamental super-selection rule which excludes the mixing of existing symmetrization groups.
In-depth discussion of this topic is beyond the scope of this work, 
and we refer the reader to the review by Okun \cite{Oku89} in 1989 and the more recent one by Elliot and coworkers \cite{Ell12} for a list of experiments and a comparison of past limits on the probability of PEP violation. 

\medskip

The VIP-2 experiment (VIolation of Pauli exclusion principle - 2) is a greatly improved version of the concept pioneered by Ramberg and Snow. 
In the context discussed in the previous paragraph, we limit the discussion and comparison of the $\frac12\beta^2$ parameter only to this type of measurement.
We have published a preliminary result from the first data-taking campaign in a brief report \cite{Cur17}, 
and in this paper we present a full account of the experiment, including the details of the analysis which introduced improvements in the systematics comparing to the preliminary result, 
and the details of the improved Monte Carlo simulation for the detection efficiency factor. 
The structure of the paper goes as follows.
In section 2 we review the experimental method and introduce the setup of the VIP-2 experiment. 
The analysis and the result of the data from the three-month data taking campaign in 2016 are discussed in section$~$3, 
including details of a dedicated Monte Carlo simulation for the full setup. 
We conclude the paper with future perspectives of the experiment, 
and some comments related to the future development on the interpretation of the experimental result.
In the appendix, we present the outline of the calculation for the expected $2p$-$1s$ transition energy that violates the PEP for copper atom, 
with further details presented in the technical note of LNF-INFN in 2013 \cite{Cur13}.

\begin{figure*}[htbp]
\centering
\includegraphics[width=15cm,clip]{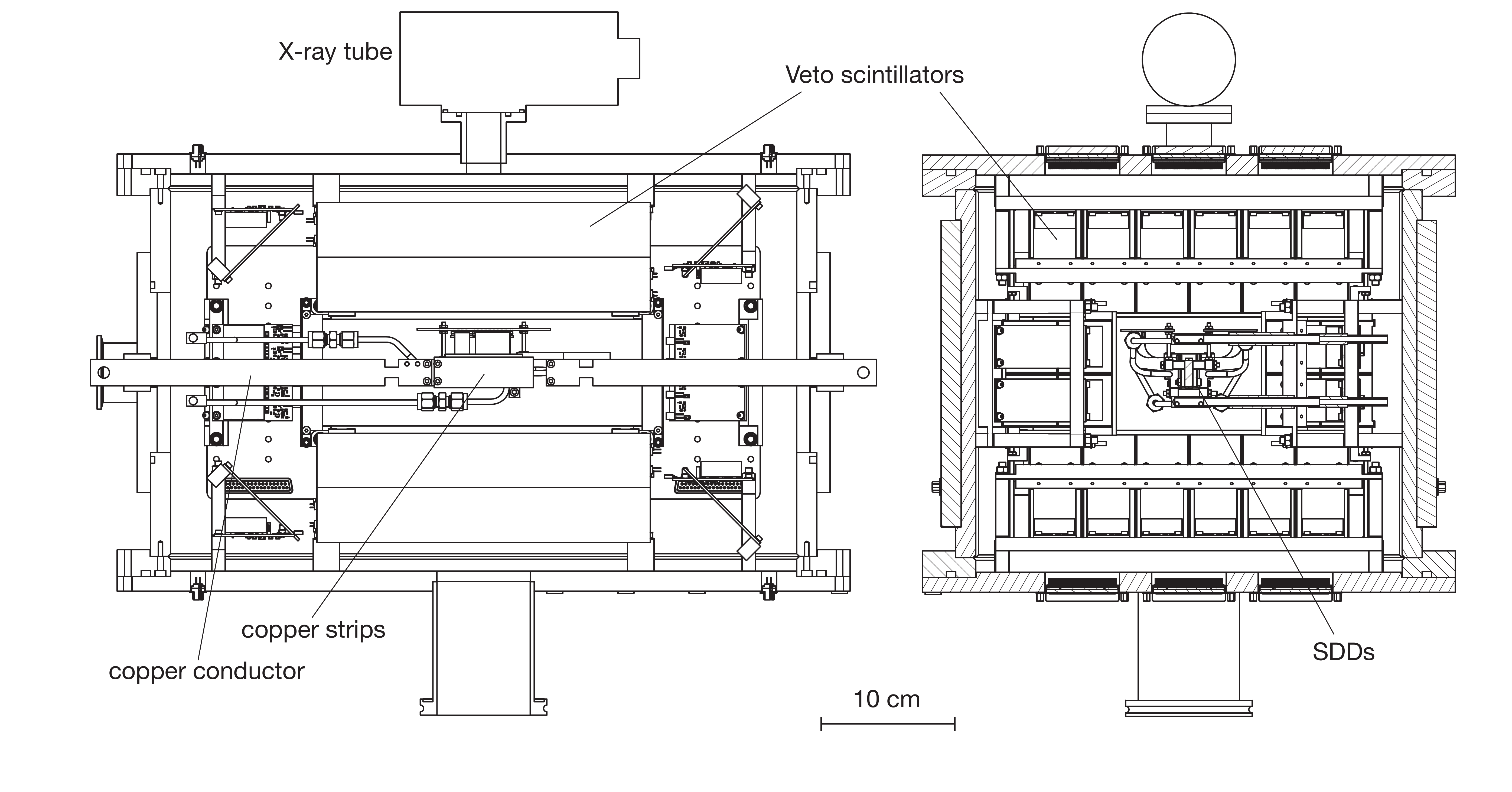}
\caption{ The side views of the design of the core components of the VIP-2 setup, including the SDDs as the x-ray detector,
          the scintillators as active shielding with silicon photomultiplier readout.
}
\label{fig:setup} 
\end{figure*}

\begin{figure*}[htbp]
\centering
\includegraphics[width=17cm,clip]{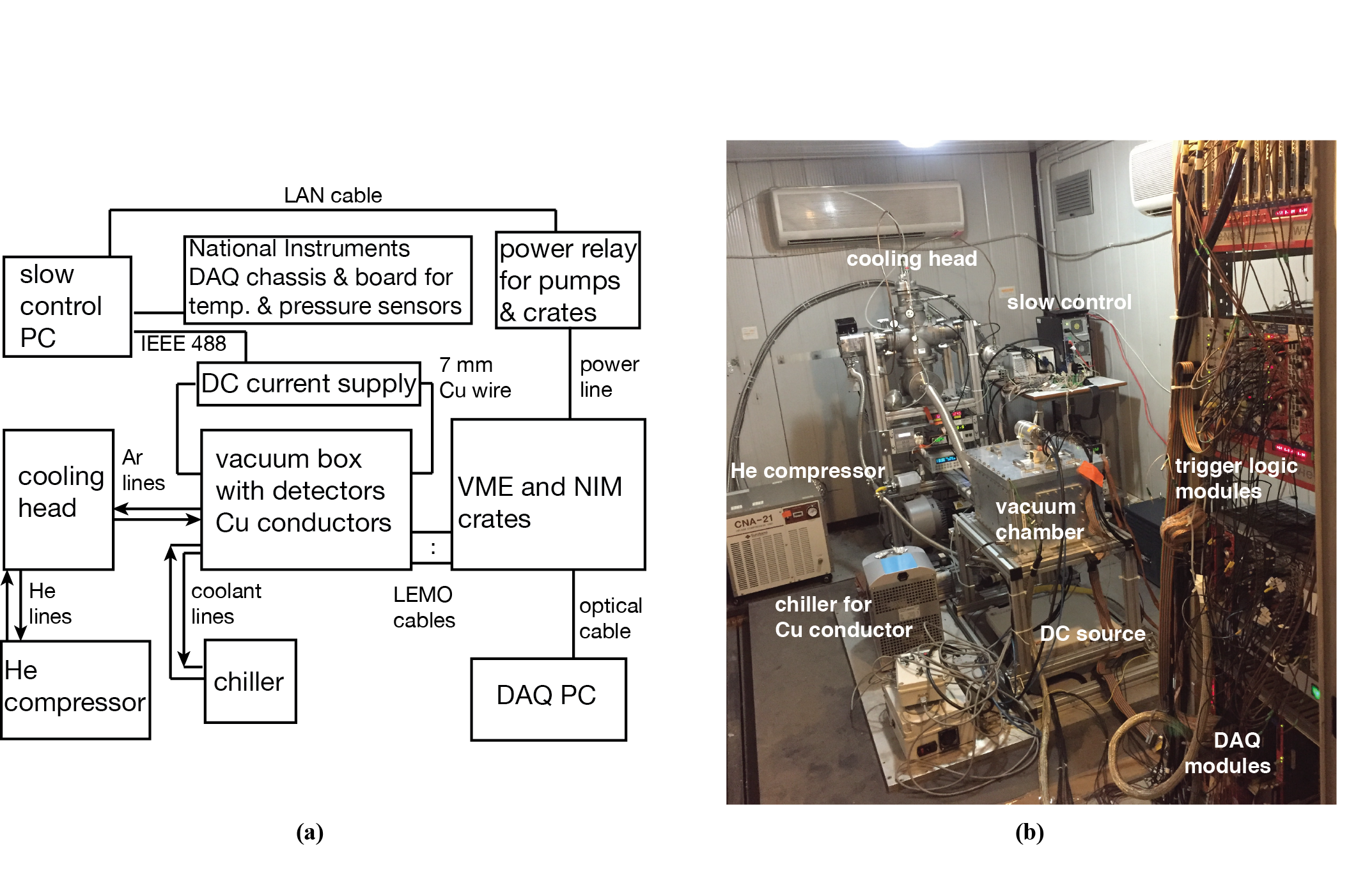}
\caption{(\textbf{a}) The diagram of the components of the experiment and their configuration; 
         (\textbf{b}) a picture of the VIP-2 setup in operation at the underground laboratory of Gran Sasso.
}
\label{fig:barrack}
\end{figure*}

\section{VIP-2 experiment}
The VIP-2 experiment is an upgrade of the VIP experiment at the Gran Sasso Laboratory, 
which had set the best limit 
of PEP violation using the experimental method of Ramberg and Snow. 
By performing the measurement at the LNGS underground low radioactivity laboratory for low rate of cosmic ray origin background, 
the VIP experiment further made use of the Charge Coupled Device (CCD) as the x-ray detector with a typical energy resolution of 320$~$eV at 8$~$keV which 
allowed a precise definition of region of interest within which the anomalous x-rays are expected to be observed.
Using the same parameter definition introduced by Ramberg and Snow for the probability that the PEP is violated,  
the VIP experiment set a limit for the probability of the PEP violation for electrons of 4.7$~\times~10^{-29}$ \cite{Cur11} \cite{Bar09} \cite{Spe08}.
In the VIP-2 experiment, we intend to further improve the sensitivity by two orders of magnitude in the search for PEP violating events, 
with the application of Silicon Drift Detector (SDD) \cite{Fio06} \cite{Lec01} as x-ray detectors with better energy resolution.
The timing capability of the SDDs also enables us to introduce an active shielding made of plastic scintillators, as illustrated in Figure \ref{fig:setup} (a), 
which removes the background originating from the high energy charged particles that are not shielded by the rocks of the Gran Sasso mountains.
The dominant background from the environmental radiations will be shielded by the passive shielding made of lead and copper blocks in the similar way as in VIP experiment. 
Major improvements also come from the new layout of the strip shape of the copper conductor, which allows a larger acceptance for the x-ray detection by the SDDs. 
A DC current of 100$~$A instead of 40$~$A introduces more than twice as many new electrons into the copper strip, thus increases the candidate event pool for the anomalous x-rays. 
Details of the instrument designs and the detector performance tests of the VIP-2 setup are given in \cite{Shi16} \cite{Pic16} \cite{Shi15} \cite{Mar13}. 

We started the first data taking campaign for VIP-2 experiment in 2016, 
after the setup was transported and mounted in the Gran Sasso underground laboratory as shown in Figure$~$\ref{fig:barrack} (b).
The physics data taking was preceded by a series of tests for the detector performance and tuning of the apparatus, 
both at the above ground laboratories and at LNGS.
An in-situ energy calibration scheme for the SDDs was implemented for the physics run, 
and the scheme constantly provided reference energy peaks to calibrate the digitized SDD signals to energy scale. 


\subsection{Experimental setup}
The central part of the experiment consists of a copper conductor in the shape of strips, 
two arrays of 1$\times$3 Silicon Drift Detectors (SDDs) on each side of the copper strip, each array with 3$~$cm$^2$ of effective surface,
and 32 pieces of plastic scintillators as active shielding surrounding the SDDs, to veto the background originating from the cosmic ray events. 
All the detectors and the front end preamplifier electronics are mounted inside the vacuum chamber shown in Figure$~$\ref{fig:setup} kept at 10$^{-5}~$mbar during operation. 

Each of the two strips of copper has a thickness of 50$~\mu$m, and a surface of 9$~$cm$~\times~$2$~$cm.
They are coupled to two thick copper conducting bars, which go through the vacuum chamber, and the external ends of the bards are connected to a DC current supply.
The heat due to the dissipation in copper leads to a significant temperature rise in the strip, 
so a cooling pad is placed in between the two strips and cooled by a closed circuit of the Fryka chiller. 
When a DC current of 100$~$A is applied to the copper conductor, the temperature rises up by about 20$~^{\circ}$C. 
With this cooling configuration, the temperature rise at the SDDs was about 1$~$K when the current was on. 
At this level of temperature difference, the performance of the SDDs has no significant change for our measurement.

The SDDs are cooled down to -$~$170$~^{\circ}$C with circulating liquid argon in a closed cooling line, 
and the argon is cooled with a Sumitomo CNA-21 helium compressor. 
Front end preamplifiers are connected within few centimeters from the SDDs, 
and the analog outputs are sent via LEMO coaxial cables through the feedthroughs on the vacuum chamber to be further processed for trigger and digitization. 

\begin{figure*}[htbp]
\centering
\includegraphics[width=16cm,clip]{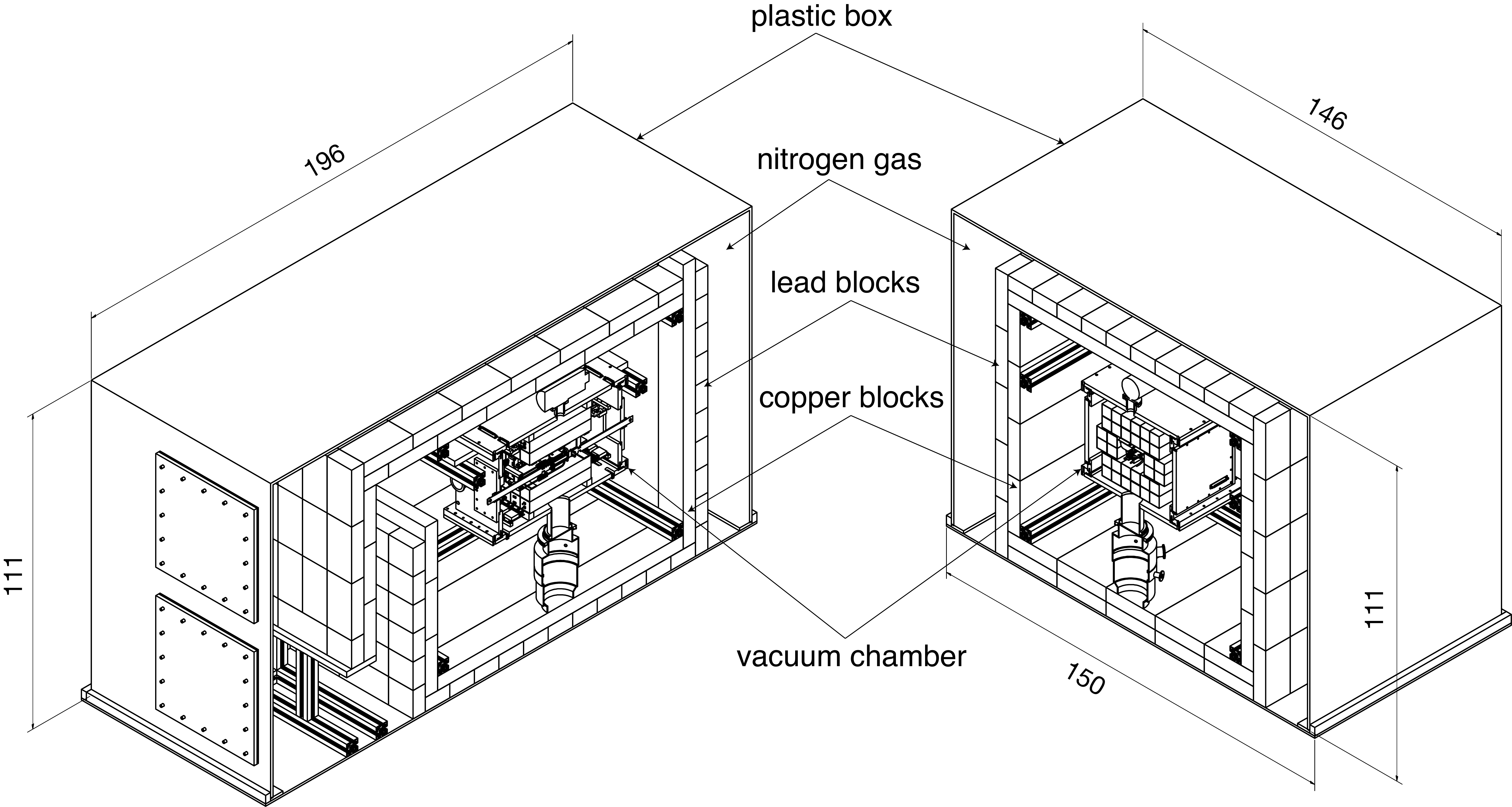}
\caption{ 
          Perspective views of the VIP-2 apparatus with passive shielding, with the dimensions in cm.
          Nitrogen gas with a slight over pressure with respect to the outside air will be circulated inside the plastic shielding. 
}
\label{fig:shielding}
\end{figure*}

As active shielding surrounding the SDDs, we use 32 pieces of plastic scintillators each in the shape of a \\
250$~$mm$~\times~$38$~$mm$~\times~$40$~$mm bar.
This veto detector will identify the few tens of cosmic ray events passing through the setup each day, 
which are going to be a dominant background source when the fully implemented passive shielding as shown in Figure$~$\ref{fig:shielding} shields out the environmental gamma radiations.
To read out the light output of each scintillator bar, two pieces of silicon photomultipliers (SiPMs) are coupled to one end of each bar.
The amplification electronics for the SiPMs are also mounted inside the vacuum chamber, 
and the analog outputs are sent out in similar way as those for the SDDs. 

In order to make quick check for the function of the SDDs, 
the top of the vacuum chamber has a kapton window, 
and some scintillator bars above the SDDs have an opening solid angle to let the x-ray tube on top of the setup irradiate zirconium and titanium foils to produce fluorescence reference lines. 
With this quick energy calibration, in one hour we can collect data with enough statistics to determine the resolution and the performance of the SDDs. 

The complete configuration of the apparatus in the VIP experiment barrack at the Gran Sasso Laboratory is illustrated in Figure$~$\ref{fig:barrack}. 
As the DC current source to the copper conductor, we used a commercial Agilent power supply module N5761A, 
which has a maximum of current up to 180$~$A. 
The module was controlled remotely, integrated with the slow control system.

\begin{figure*}[htbp]
\centering
\includegraphics[width=15cm,clip]{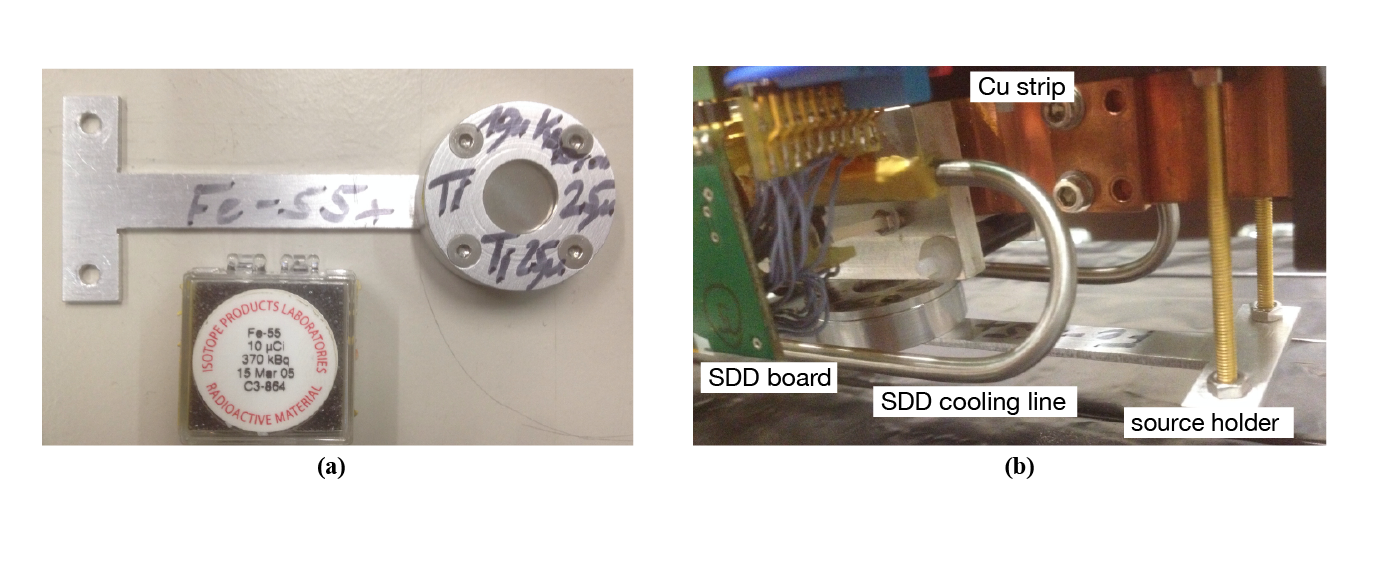}
\caption{(\textbf{a}) The holder of the source with the Fe-55 source inside covered by a 25$~\mu$m thick high purity titanium foil; 
         (\textbf{b}) a photo of the holder mounted near the copper strip and the SDDs. 
}
\label{fig:source}
\end{figure*}

In addition to the x-ray tube, we have a secondary energy calibration method of the SDDs using a weakly radioactive Fe-55 source.
A 25$~\mu$m thick titanium foil was attached on top of the source and then mounted together inside an aluminum holder, 
as illustrated in Figure$~$\ref{fig:source}.
With this configuration, the six SDDs have an overall 2$~$Hz trigger rate, 
accumulating events of fluorescence x-rays from titanium and manganese to calibrate the digitized channel into energy scale. 
Since the calibration energies are below the energy region of the candidate events, 
the source is kept inside the setup throughout the data acquisition thus realizing an in-situ energy calibration.

\begin{figure*}[htbp]
\centering
\includegraphics[width=15cm,clip]{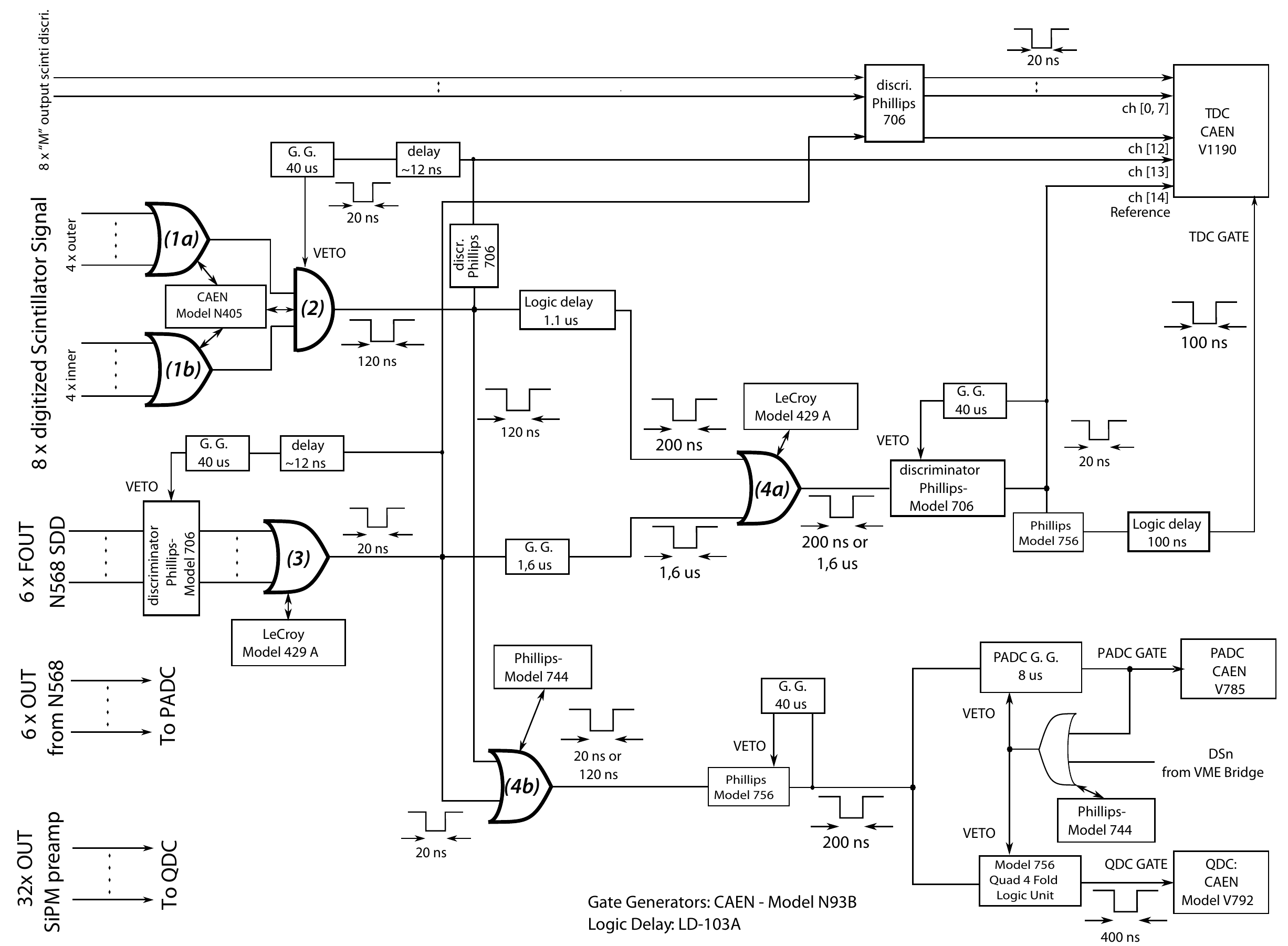}
\caption{The diagram of the trigger logic. The essential parts are highlighted with explanations in the article. 
}
\label{fig:trigger}
\end{figure*}   

\subsection{Trigger logic}
The trigger logic was constructed with the standard nuclear instrumentation modules (NIM), 
with the explicit diagram shown in Figure$~$\ref{fig:trigger}.
Key components that define the events at the veto scintillators and the SDDs are highlighted and the meanings are as follows. 
The 32 pieces of veto scintillators surround the SDDs in two layers, 
and an event at the veto is defined as : any hit at each layer, as marked with (1a) for outer layer and (1b) for inner layer in the diagram, 
and then the AND coincidence of the two layers marked as (2) defines the veto event. 

One SDD event is simply given by any event at one of the six SDDs, marked as the OR logic (3) in the diagram. 
Then the OR logic (4b) between the veto scintillator event (2) and the SDD event (3) gives the main trigger for data taking of charge-to-digital-converter (QDC) 
and peak-value-analog-to-digital-converter (PADC). 
On the other hand, because the SDD signal has a time scale at the order of microseconds comparing to the nanosecond-order time scale of the scintillators, 
some adjustment of timing was further implemented at (4a) to provide the time gate for the time-to-digital-converter (TDC) module.

\subsection{Data acquisition}
The data taking system is based on the VME standard interface. 
A controller module communicates with the TDC, QDC, and PADC modules, 
and whenever there is one event ready at the register of the modules, 
the data is transferred to the acquisition computer without buffering more data in the register of the modules.
No triggers are accepted during the communication and data transfer, 
and a maximum trigger rate of about 150$~$Hz can be accepted with this acquisition mode. 

Each event includes the following information : 
\begin{itemize}
\item  trigger rate at the time of the event; 
\item  QDC of 32 channels SiPM readout of scintillators; 
\item  PADC of the digitized SDD signals from the CAEN N568 spectroscopic shaping amplifier;
\item  timing of the SDD event and the coincidence of the veto scintillators w.r.t. the trigger;
\item  time tag of the event obtained from the clock of the data taking computer. 
\end{itemize}

It records the energy deposit of the six SDDs, from the output of a CAEN$~$N568 spectroscopy amplifier which processes the analog signals of the SDD preamplifier output. 
The QDC signals of the 32 scintillator channels, and the timing information of the SDDs with respect to the main trigger are recorded in the data. 
The data acquisition system can be remotely accessed and controlled from the computer terminals outside the Gran Sasso laboratory, 
and the real time spectra from all the modules can be monitored from an online display.

\subsection{Slow control}
A slow control system monitors online and records the temperatures of the SDDs, the copper conductor, the cooling system, and the ambient temperature, as well as the vacuum pressure of the setup. 
The system which can be accessed from remote terminals controls the DC power supply to switch on and off the current applied to the copper strip, 
and also the SDD power supply and the turbo pumps for the vacuum. 
An automatic safety procedure was implemented to shut down the power supplies in case of a gas leak or temperature rise of the devices. 

A closed circuit chiller coupled to a cooling pad attached to the copper strips keeps a constant temperature below 25$~^{\circ}$C of the strips 
when the DC current up to 100 A is applied. 
The temperature of the SDDs' holder frame had a change of less than 1 K when the 100 A current is applied to the copper strip. 
At this level of temperature variation, the effect of change in the energy resolution of the SDDs is negligible.

\subsection{Preliminary tests}
A series of tests for the basic performance of the detectors and the setup were done at the Beam Test Facility (BTF) in Frascati and at the Stefan-Meyer-Institute (SMI) in Vienna. 
Using the electron and positron test beam at BTF, we measured the efficiency and the response of the scintillator with SiPM readout to the charged particles. 
The confirmed efficiency of above 97\% and the time response of 2$~$ns \cite{Shi15} were sufficient to identify cosmic ray events that lead to background at the SDDs. 
With the detectors and the trigger logic assembled at SMI, we tested the veto scintillators together with the SDDs. 
Using the cosmic ray events, we confirmed the timing response of the SDDs at 400$~$ns FWHM \cite{Shi16}. 
Cooling test for the copper strip was also performed. 
With a DC current up to 180$~$A applied to the copper conductor, the copper strip temperature stayed at the room temperature level below 25$~^{\circ}$C. 
The data taking system and the trigger logic were fine tuned using the cosmic ray events. 
The energy calibration configuration for the SDDs was optimized, and all the six SDDs achieved the energy resolution better than 150$~$eV FWHM at 6$~$keV. 
During the laboratory test, we also confirmed the vacuum and the argon cooling system stability, the functionality of the slow control and the safety procedure.

In November 2015, after the exhaustive tests,  
the VIP-2 setup was transported and mounted in the Gran Sasso underground laboratory as shown in Figure$~$\ref{fig:barrack} (b).
After tuning and optimization, from October 2016 we started the first campaign of data taking with the complete detector system. 
A total amount of 34$~$days of data with a 100$~$A DC current and 28$~$days without current were collected until the end of the year 2016. 
In the next section, the analysis and the result of this data set is discussed in detail.


\begin{figure}[htbp]
\includegraphics[width=9.5cm,clip]{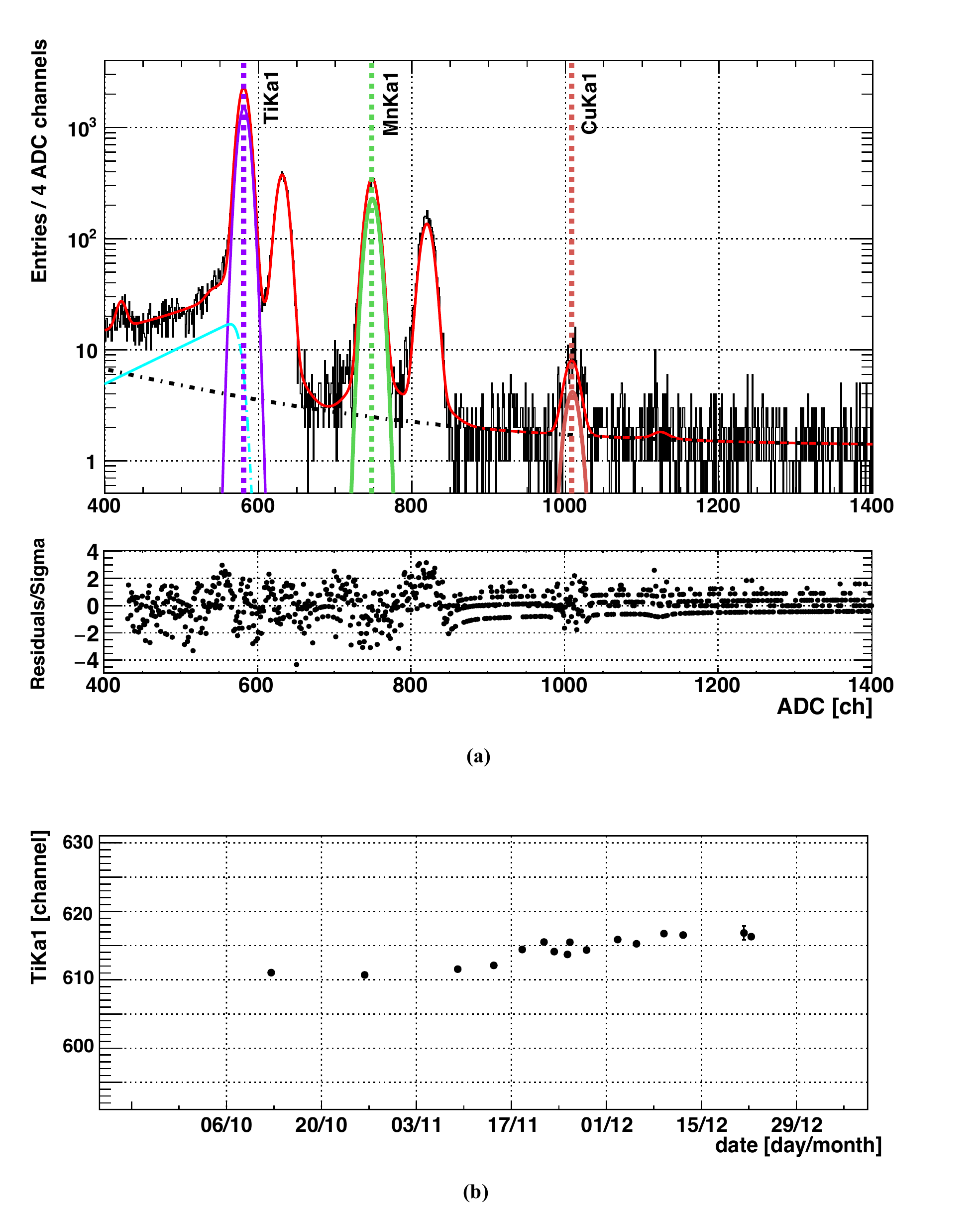}
\caption{(a) A typical fit to the PADC data from one SDD for one week;
         (b) the peak position of the titanium $K_{\alpha1}$ from calibration fit to each data subset over the full data taking period, 
         note that most of the statistical errors are smaller than the radius of the plot dots. 
}
\label{fig:calib}
\end{figure}

\section{Data analysis}
The main objective of the 2016 data analysis is in the energy spectra of the SDDs. 
Without a passive shielding, the dominant background at the SDDs comes from the environmental gamma radiations inside the barrack. 
This fact is discussed in the section on Monte Carlo simulation.
Since the veto scintillators were designed to be sensitive to cosmic rays with energy deposit larger than few MeV, 
they were not used for the background rejection in this data set. 
However they will become effective with the final configuration of the VIP-2 setup, 
which will have a passive shielding to remove the gamma background when the cosmic ray origin background will become dominant. 
From the comparison between the QDC data of the scintillators taken at SMI and LNGS, 
we confirmed preliminarily that a reduction rate of the cosmic ray event is at least at the order of 10$^{-4}$.

\begin{figure*}[htbp]
\centering
\includegraphics[width=12cm,clip]{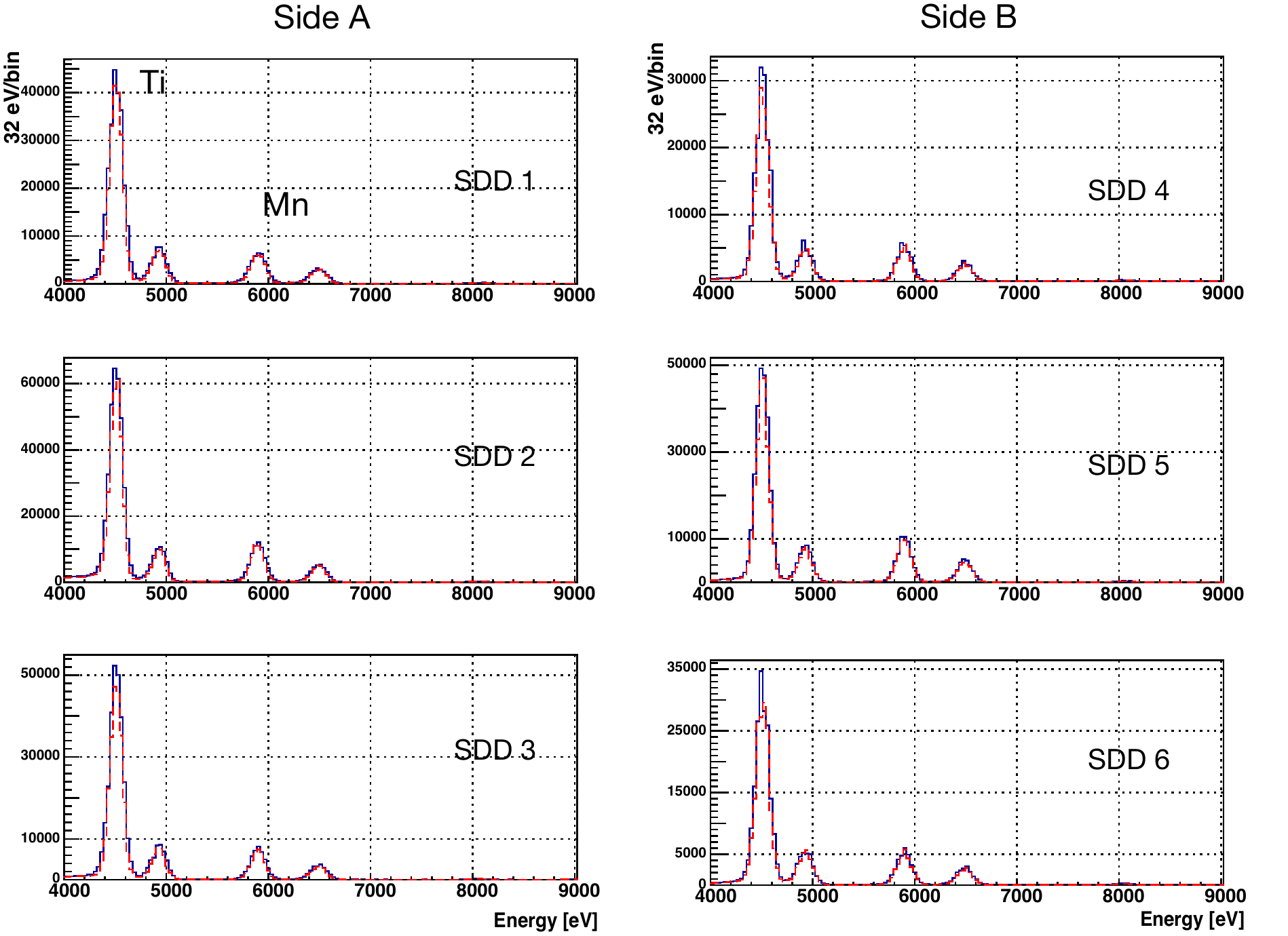}
\caption{The energy spectra of six individual SDD units; 
         data taking with current : dashed lines; data taking without current : solid lines. Copper lines are not visible with a linear vertical axis scale. 
         All spectra are in their full statistics without normalization with exposure time.
}
\label{fig:spectra}
\end{figure*}

\subsection{SDD energy calibration}
During the 2016 data taking campaign, the DC current was switched on typically for one week and off for the next. 
The energy calibrations for the SDDs were done for each data subset corresponding to a period of about one week. 
In the fit to each ADC spectrum from one week of one SDD, 
the range was chosen from about 3$~$keV to above $10~{\mathrm  k}$eV that covers escape peak of the titanium $K_{\alpha}$ peaks and the copper lines near 8$~$keV. 
For the titanium, manganese, and copper fluorescence x-rays, 
the $K_{\alpha1}$, $K_{\alpha2}$, and $K_{\beta}$ lines were each presented by a Gaussian function, 
plus an exponential tail function near the lower energy part of the Gaussian to empirically represent the skewed experimental line-shape.
This fit function has been well established in former experiments \cite{Oka07} \cite{Sidt09} where the same type of SDDs were used in similar conditions. 
Then by comparing the channel positions of the titanium and manganese $K_{\alpha1}$ peaks to their reference energy values, 
we determine the correspondence between the channel and photon energy and obtain the spectrum with energy scale in electron-volt, 
to be summed with other data subsets of all SDDs. 
From (b) in Figure$~$\ref{fig:calib} one can see that over the long data taking period the peak positions can drift from 5 to 10 ADC channels. 
The in-situ calibration to the data subsets reduced this instability over time, 
which is demonstrated by the fact that the spectrum of the summation of all SDD data has a compatible energy resolution as obtained from one typical calibration fit.

The spectra for each SDD that correspond to 34$~$days of effective data acquisition with 100 A current on and 28$~$days with current off are shown in Figure$~$\ref{fig:spectra}, 
in which the fluorescence lines of titanium and manganese are marked, 
and the statistics of the copper fluorescence lines are not visible in the linear vertical scale plot. 
We confirmed that the statistics are at the similar level for all the SDDs with similar behavior. 
We only found that the side ``A'' has slightly more statistics due to the source holder that was not perfectly aligned. 
Finally all the data subsets are summed separately over the current-on and off runs, 
ready for the simultaneous fit to be introduced in the next section.

\begin{figure*}[htbp]
\centering
\includegraphics[width=17.5cm,clip]{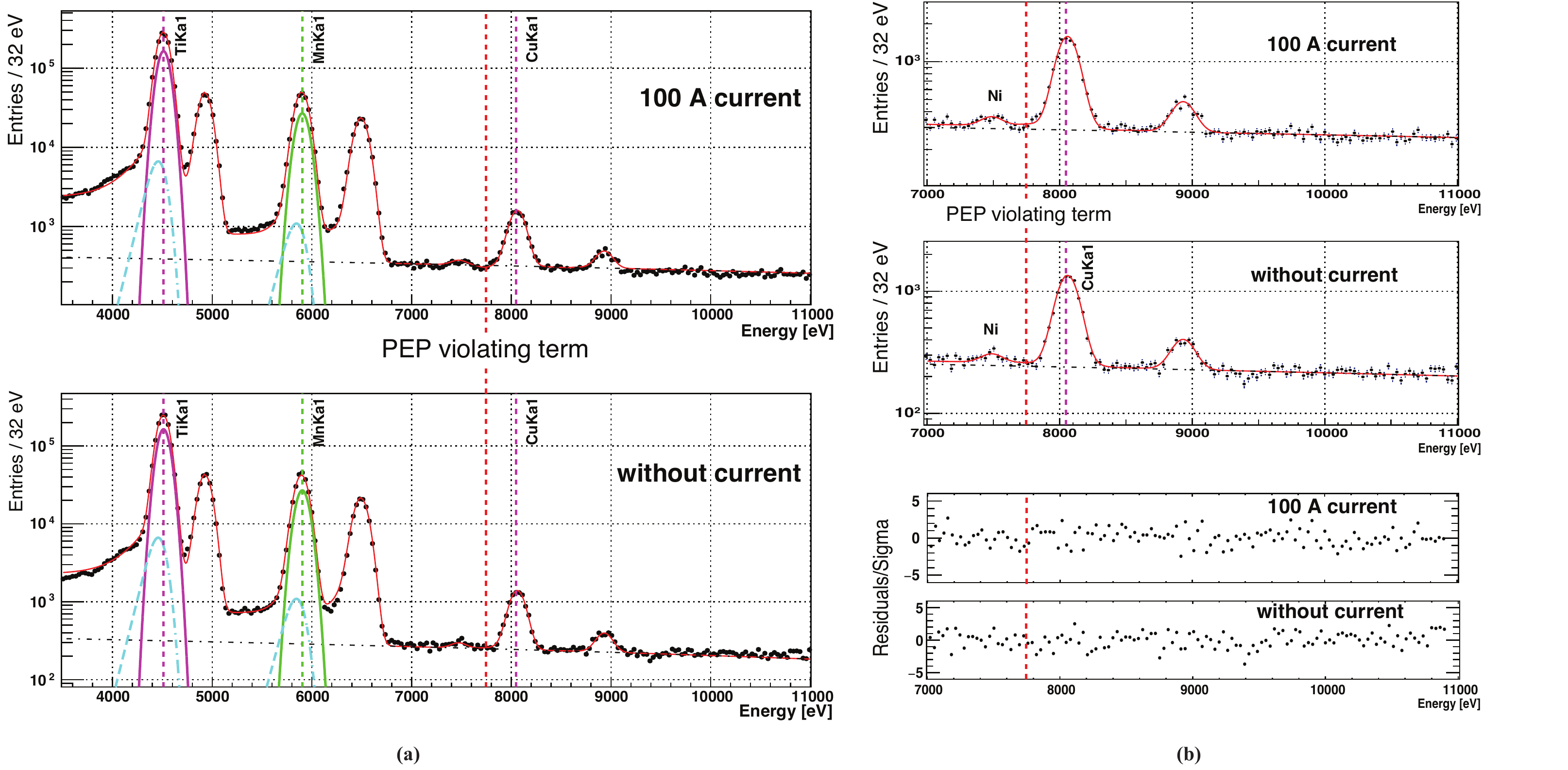}
\caption{A global chi-square function was used to fit simultaneously the spectra with and without 100$~$A current applied to the copper conductor. 
         The energy position for the expected PEP violating events is about 300 eV below the normal copper $K_{\alpha 1}$ transition. 
         The Gaussian function and the tail part of the $K_{\alpha1}$ components and the continuous background from the fit result are also plotted. 
         (a) : the fit to the wide energy range from $3.5~{\mathrm k}$eV to 11 keV; (b) : the fit and its residual for the 7 keV to 11 keV range where there is no background coming from the calibration source. 
         See the main text for details.
}
\label{fig:simfit}
\end{figure*}

\subsection{Simultaneous fit}
We performed a simultaneous fit to both current on and off spectra to determine in the current on spectrum the component that corresponds to the PEP violating candidate events. 
A similar method has been applied to the analysis of SDD spectra in an experiment of kaonic atom x-ray precision spectroscopy \cite{Sidt11} \cite{Sidt16}.
In the analysis, a global Chi-square function was defined for the two histograms of energy spectra. 
From a statistical point of view, this corresponds to the construction of a single likelihood function by taking the products of the individual likelihood functions 
of the two different experimental conditions, and assuming a Gaussian statistics for the measurement errors. 
The parameters that represent the detector energy resolution, the shape of the continuous background, 
the shape of the fluorescence peaks, are common for both spectra. 
On the other hand, the parameters for the intensities of the fluorescence peaks and the continuous background, 
are separately defined for current on and current off spectra. 
In addition, in the function for the current-on spectrum, one Gaussian component was included to represent the candidate PEP violating events. 
The center of this Gaussian component is set to be at 7746.73 eV, about 300 eV below the $K_{\alpha1}$ peak of the copper, referring to \cite{Cur13} and the Appendix in the end of this article, 
and the intensity corresponds to the number of events, which is set to be a free parameter.
With this method, we can obtain the number of candidate events and its statistical error directly from the result of a converged minimum chi-square fit, 
in this case we used the MINUIT package of the CERN ROOT software framework \cite{Root}.  

\begin{table*}[h]
\caption{ Free parameters for the simultaneous fit function. Common parameters are placed in the center of the two middle columns.}
\centering
\label{tab:fitpar}
\begin{tabular}[]{l | c c | c}
\hline
Parameters      &     100 A current    &  no current    &     Description    \\
\hline 
\multirow{2}{*}[0cm]{ First order polynomial function }      & $pol0_{WC}$   &   $pol0_{NC}$    &   \multirow{2}{*}[0cm]{ Continuous background}   \\
                &   \multicolumn{2}{c|}{$pol1$}              &                      \\ \hline
Energy-dependent &   \multicolumn{2}{c|}{$Fano$}              &    Fano factor \\ 
resolution parameters      &   \multicolumn{2}{c|}{$Constant$ $noise$}  & Energy-independent noise contribution \\ \hline
Energy scale shift  & \multicolumn{2}{c|}{$\Delta E$}        & Non-linearity at Cu \\ \hline
 \multirow{5}{*}[0cm]{ Background from fluorescence x-rays. }
                &   $CuK_{\alpha 1}A_{WC}$  &     $CuK_{\alpha 1}A_{NC}$              &  Intensities of Cu$K_{\alpha 1}$ \\
 \cline{2-3}    &   \multicolumn{2}{c|}{$CuK_{\alpha 1}eV$}     &   Mean value of Cu$K_{\alpha 1}$    \\
                &   \multicolumn{2}{c|}{$CuK_{\beta}eV$}        &   Mean value of Cu$K_{\beta}$    \\
                &   \multicolumn{2}{c|}{$CuK_{\beta}$/$K_{\alpha1}$}     &  Cu $K_{\beta}$ to $K_{\alpha1}$ relative intensity \\ 
                &   \multicolumn{2}{c|}{$NiK_{\alpha1}$/$CuK_{\alpha1}$} &  Ni $K_{\alpha 1}$ to Cu $K_{\alpha1}$ relative intensity \\ 
\hline
PEP violating component     &  $CuPEPAmp$    &             & Intensity of PEP violating component \\ 
\hline
\end{tabular}
\end{table*}

The simultaneous fit consists of two steps. 
In the first step, as shown in Figure$~$\ref{fig:simfit} (a), a wide energy range from 3.5 keV to 11 keV was chosen, 
to use the high statistics titanium and manganese fluorescence lines to determine the energy dependent resolution parameters of the detectors, 
given by the Fano Factor and the Constant Noise that represents an energy-independent contribution to the energy resolution. 
Then using the result from the wide energy range fit as the input parameter, 
a second fit was done for the energy range from 7 keV to 11 keV, where the manganese and the titanium lines were excluded. 
This allows a better determination of the shape of the continuous background at the energy range above the calibration sources.
The number of the candidate PEP violating events is obtained from the second step of the simultaneous fit, 
the result of which is plotted in Figure \ref{fig:simfit} (b), where the bottom two plots show the residuals of the fit. 

We summarize the free parameters in the second step of the fit in Table \ref{tab:fitpar}.
The continuous background is described with a first order polynomial function.
As the two data sets were taken under the same background conditions, which should lead to identical shape represented by the slope of the first order polynomial background, 
we introduced a common $pol1$ parameter for spectra both with and without current. 
Then we used two independent $pol0$ parameters to represent the different background levels of the two spectra due to the difference in exposure time. 
The ratio between these two intensities from the fit is consistent with the ratio of exposure times. 
Similar argument applies to the background from the fluorescence x-rays when the copper components are irradiated : 
the mean value of the peaks, the relative intensity between $K_{\beta}$ and $K_{\alpha1}$ lines is common for the two data sets, 
while the absolute intensities of the Cu$K_{\alpha1}$ are independent free parameters. 
In this way the statistics of the two spectra was used as it is without normalization using the exposure time, 
and any effect from the uncertainty of the exposure time is included in the fit error of the PEP violating component amplitude, 
which is the only physics parameter of the fit. 
The Fano factor and the Constant Noise that describe the energy-dependent resolution are also free parameters, 
and the energy resolutions obtained from the two steps of the fit were confirmed to be consistent in 190 eV FWHM at 8.0 keV.

Note that the copper fluorescence x-rays are the background for our candidate event, 
and they originate from the copper when irradiated by the gamma background in the barrack or the cosmic rays. 
We also included the $K$-series lines from nickel in the fit as we confirmed the existence of this component from data taken in the first half of 2017 with three times the statistics.
However we cannot yet identify the exact origin of the nickel inside the setup. It will be one of the studies to be carried out in the future. 

Compared with the method of subtracting the spectra with and without current, which has been used in previous publications where the same experimental method was used, 
the simultaneous fit has the following differences and advantages : 

\begin{itemize}
\item  normalization :  
since the principal idea of the spectra subtraction requires that the current-off data should represent the background of the current-on data, 
in order to subtract two histograms, a normalization with data taking time is usually necessary. 
This procedure introduces a systematic error that could not be precisely assigned. 
The simultaneous fit on the other hand does not need a normalization to the histograms, 
since the intensities of the fluorescence lines and the continuous background are free parameters in the fit. 
Therefore the statistics from the current on/off data can come from different amount of data taking time, and can be used directly. 

\item  Region Of Interest (ROI) : 
an energy region within which the number of events for the current on/off spectra must be defined in order for the subtraction. 
The width of the region is usually given by the FWHM of the detectors' energy resolution. 
In the subtraction method, events are counted in one bin, and the statistical error of the count is derived assuming a Poisson distribution for the number of events.
However, the simultaneous fit does not require the definition of ROI,
since a wide energy range of the spectra is considered in determining the parameters of the global function, 
from the result of which the number of candidate PEP violating events is obtained. 

\item  error assessment : 
as the number of candidate PEP violating events is defined as a fit parameter, 
the statistical error given by the fit algorithm takes into account the uncertainties of other parameters, 
which include the energy resolution, the shape of the continuous background, the amount of copper fluorescence x-rays which are inside the ROI. 
These uncertainties are usually not evaluated when using the subtraction between the signal and background spectra. 
Additionally, the present method also returns correlations between parameters, 
and from the correlations we can infer the importance of the different features of the global spectrum in the determination of the final bound on PEP violation. 
\end{itemize}

In our recently published brief report \cite{Cur17} based on the same data set, 
we obtained the number of PEP violating events from the subtraction of the signal and background spectra as $N_{X}=41\pm66$. 
We will show in the results section that the analysis method of this work achieved compatible uncertainty in the number of events. 
However we need to remark that, even though the uncertainties are compatible at the numerical level, 
the result from this work has taken into account contributions from systematics that could not be well evaluated using the spectra subtraction. 
It is important to note the improvement in the control of the systematics that is not self-evident in the numerical result.

\subsection{Monte Carlo simulation}

\begin{figure*}[htbp]
\centering
\includegraphics[width=17cm,clip]{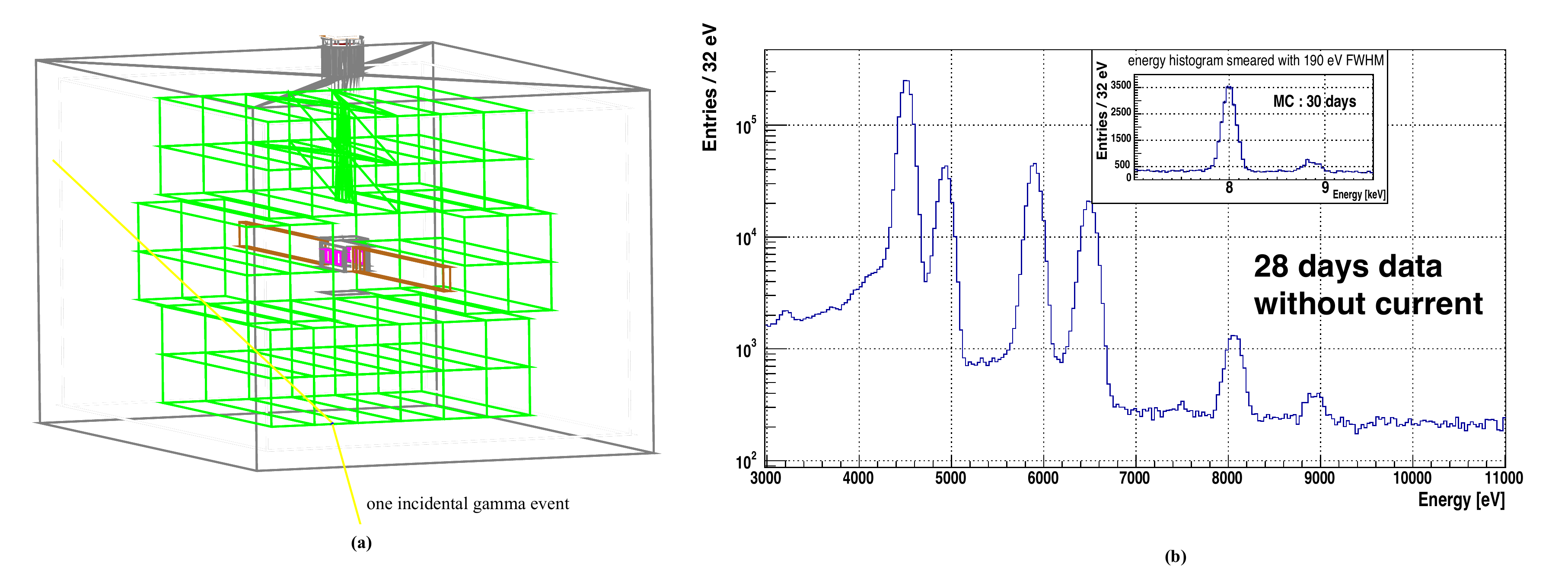}
\caption{(a) The geometry of the setup used in the Geant4 simulation;
         (b) a comparison between the 28$~$days of no current data and the simulated energy spectrum (the inset) of six SDDs from 30$~$days of data taking without current. 
         The input of the background source is the gamma radiation, 
         with the rate and energy distribution taken from published measurement inside the tunnel of the Gran Sasso laboratory.  
}
\label{fig:mc}
\end{figure*}   

A realistic setup that includes the SDDs and the scintillator detectors, the copper conductor and the aluminum vacuum chamber, 
as shown in Figure$~$\ref{fig:mc} (a), 
was implemented in the simulation, using the Geant4.10 toolkit with the Low Energy Penelope Physics package \cite{Pen01} for the interactions of low energy photons below 1$~$MeV. 
Three types of simulations were performed in the framework of this study. 

The first type is for the background originating from gamma rays inside the laboratory. 
As the input of the energy distribution and the rate of the gamma radiation, 
we took the data from the measurement of the gamma-ray flux below$~$500 keV in the underground Hall$~$A of LNGS \cite{Buc09}. 
The flux was applied uniformly to the outer surfaces of the aluminum box, 
and the energy deposit at the SDDs is smeared by a Gaussian distribution with 190$~$eV FWHM to represent the detector resolution.
By normalizing the statistics of the simulation result to 30 days of data taking, 
we obtained an estimated spectrum for six SDDs, which reproduces to leading order the data without DC current, 
as shown in Figure$~$\ref{fig:mc} (b).

Secondly, to estimate the efficiency of the scintillator veto system, 
we checked the detector responses to the 270 GeV/c muons as typical cosmic ray events \cite{Mac03}. 
The result shows that by applying an energy cut of 100$~$keV for the energy deposit inside the scintillator, 
with the trigger coincidence defined in the trigger logic section, we can reject 99\% of the cosmic ray background at the SDDs. 
This result will be a future guide line when the passive shielding is implemented and the cosmic ray background becomes dominant for the measurement.

The last type of simulation was done to determine the effective detection efficiency of the fluorescence x-rays near 8$~$keV for the SDDs. 
For this purpose, 8$~$keV photons are generated at a random position inside the copper strip with random initial direction. 
The photon attenuation inside the strip, the photon interactions with the materials inside the setup, 
the geometry factor of the SDDs, and the Compton effect inside the SDD's silicon volume are all taken into account. 
As a result, the detection efficiency factor, which is used in the calculation of the PEP probability in the later section, 
was estimated to be about 1.8\%.

\subsection{Background sources}
The Monte Carlo simulation result as shown in the inset of Figure \ref{fig:mc} (b), 
estimates that for a 200 eV region of the flat part below the copper flurescence x-ray component, 
the environmental gamma radiations contribute to roughly 75 background events per day. 
From the unbinned data, in the 200 eV energy region centered at the expected PEP violating copper $K_{\alpha1}$ at 7746.73 eV, 
we counted 2002 events in the 34 days of data taking with applied current and 1616 events in the 28 days without current, 
which correspond to 59 and 58 events per day for this energy region of interest. 
Although the Monte Carlo model needs to be refined to account for the overestimate of the background rate, 
it is clearly confirmed that the environmental gamma radiation is the dominant background source. 

Based on the experience of the VIP experiment that the passive shielding will reduce such background by at least a factor of ten, 
the background due to environmental gamma radiations with the complete shielding is expected to be less than 5 events per day in the energy region of interest. 
Then the cosmic ray events arriving at some ten times per day become the next dominant background source, 
with its contribution to the energy of interest still to be studied based on future data with well-calibrated veto detectors and refined Monte Carlo model. 
Further studies into more rare background sources including those introduced by the radiative isotopes inside the setup materials will follow, 
after we have achieved the expected background reduction with the full shielding and have a thorough understanding of the aforementioned two dominant background sources.

\subsection{Result of $\beta^2/2$}

From the result of the simultaneous fit, we obtained the number of PEP violating events which is compatible with zero : 
\begin{equation}
N_X = 54 \pm 67 ~ (statistical),  \nonumber
\end{equation}
where the statistical error $\Delta N_{X}~=~67$ is the standard deviation of $N_X$ obtained from the fit. 
Ramberg and Snow estimated a lower bound for $N_X$ in their original paper \cite{Ram90}: 
\begin{equation}
N_X \ge \frac{1}{2} \beta^2 \cdot N_\mathrm{new} \cdot \frac{1}{10} \cdot N_\mathrm{int} \cdot (\mathrm{geometric\; factor})
\label{eq:2}
\end{equation}
where $N_\mathrm{new}$ is the total number of ``new'' electrons injected into the system, 
the factor 1/10 is an estimate of the capture probability (per electron-atom scattering) into the $2p$ state (as in \cite{Bar06}), 
$N_\mathrm{int}$ is the minimum number of electron-atom scatterings in the electron transport from its entrance to its exit 
in the region of the copper strip that corresponds to the surface dimension of their x-ray detector,
and the geometric factor takes into account both the solid angle covered by the detector and the x-ray attenuation inside in the copper strip. 
In the VIP-2 configuration, we denote the number of new electrons with $N_\mathrm{new} = (1/e)\sum I \Delta t$ 
(to indicate that the number of new electrons is given by the current integrated over the total measurement time in units of elementary charge), 
and use the estimate $N_\mathrm{int} > D/\mu$ where $D$ is the effective length of the copper strip where the current flows through a 50 $\mu$m $\times$ 2 cm cross section, 
with $\mu$ the scattering length for conduction electrons in the copper strip, 
we can also write
\begin{equation}
N_X \ge \beta^2 \frac{D \sum I \Delta t}{e\mu} \cdot \frac{1}{20} \cdot (\mathrm{detection\; efficiency\; factor}), 
\end{equation}
where the detection efficiency factor is a more accurate estimate of the geometric factor.

We obtained the detection efficiency factor from the Monte Carlo simulation as discussed in the previous section, 
and it takes into account : 
the transmission rate of a copper $K_{\alpha}$ x-ray that origins at a random position inside the copper strip and reaches the surface; 
the geometrical acceptance of the photons coming from the surface of the copper strip arriving at the six SDD detectors; 
the detection efficiency of a copper $K_{\alpha}$ x-ray by the 450$~\mu$m thick SDD unit, 
and the value is estimated to be about 1.8\%. \\

With $\mu=3.9\times10^{-6}~$cm, $e=1.602\times10^{-19}~$C, $I=100~$A, 
and the effective length of the copper strip $D=7.1~$cm (same as used in the MC simulation), 
using the three sigma upper bound of $\Delta N_X=\pm~67$ to give a 99.7\%$~$C.L., 
we get an upper limit for the $\beta^2$/2 parameter: 
\begin{equation}
\frac{\beta^2}{2} \le \frac{3\times 67}{6.0\times 10^{30}} = 3.4 \times 10^{-29}.
\end{equation}

\begin{figure}[htbp]
\centering
\includegraphics[width=9 cm,clip]{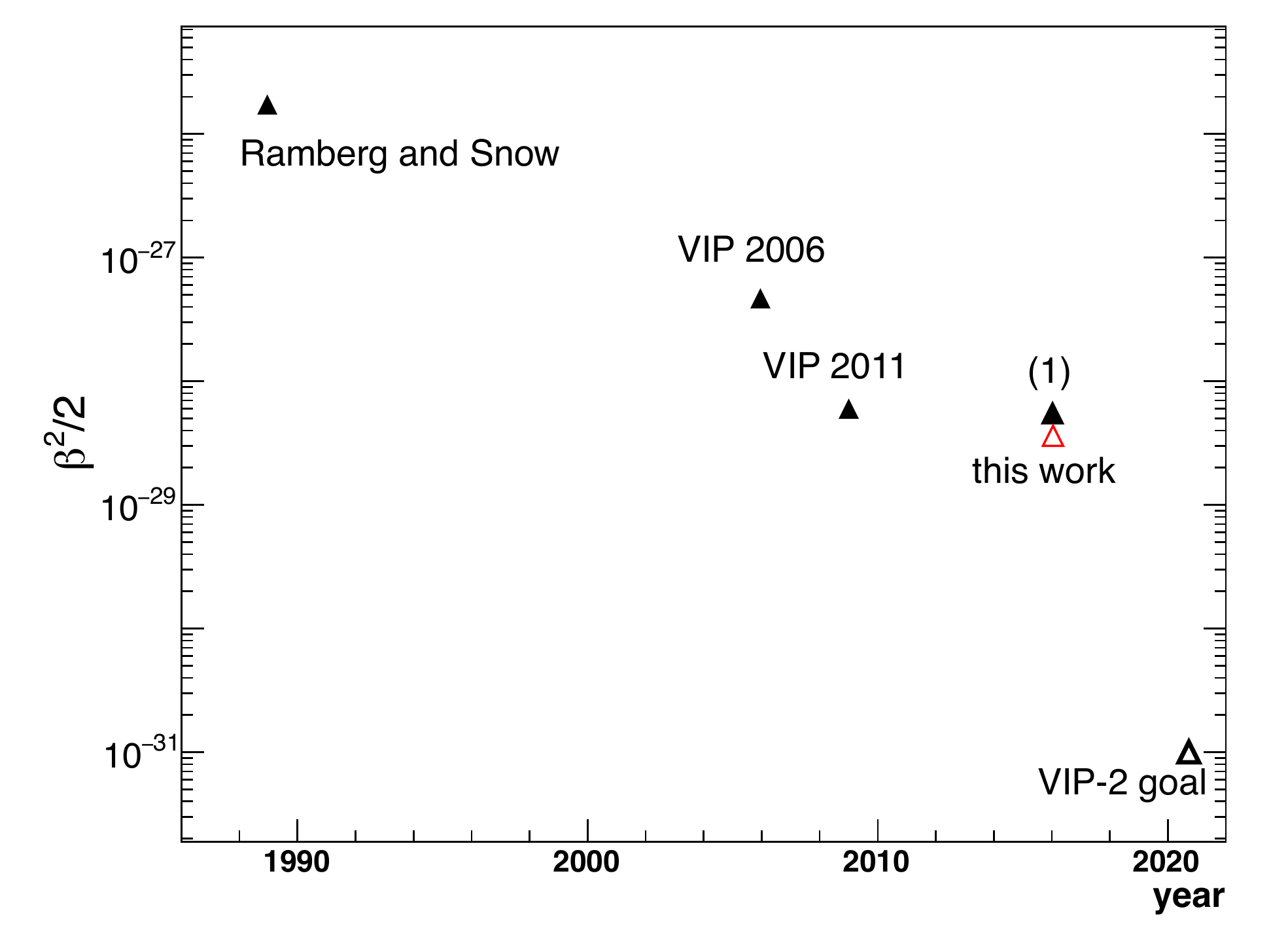}
\caption{ Past results from PEP violation tests for electrons with a copper conductor, 
          together with the result from this work and the anticipated goal of VIP-2 experiment. 
          The result (1) is based on the same data set of this work, but using the spectra subtraction in the analysis. 
}
\label{fig:betaSquare}
\end{figure}

\section{Discussions and future perspectives}
We obtained a new upper limit of the probability that the Pauli Exclusion Principle can be violated, 
from the first three months of data taking in the VIP-2 experiment. 
As shown in Figure$~$\ref{fig:betaSquare}, this work has achieved the best result among all the past experimental tests for electrons with a copper conductor. 
We introduced for the first time into the analysis the simultaneous fit of the background spectrum and the signal spectrum. 
This method eliminated the uncertainties in the normalization of the spectra and the choice of the ROI, 
which introduce systematic errors in the spectra subtraction method that were never evaluated quantitatively in previous experiments. 
With improved control of the systematic error, 
the upper limit determined from the new analysis is numerically compatible with our previously published result \cite{Cur17} marked as (1) in Figure$~$\ref{fig:betaSquare}, 
which was derived from the spectra subtraction based on the same data set of this work. 
The improvement compared to (1) came from more accurate Monte Carlo simulation for the detection efficiency factor.
(In (1), we used a detection efficiency factor of 1\% which was estimated at the design stage of the experiment, using 9 cm as the length of the copper strips.)

The Ramberg and Snow formula Eq. \ref{eq:2} is very conservative and returns a rather weak upper bound. 
A thorough review including comparison of the different experimental methods and new interpretations of the Ramberg-Snow type measurement requires an in-depth investigation of the atomic and solid-state physics models of the interactions of free electrons in metal, 
and we plan to dedicate a forthcoming paper to this topic. 
In the present article, for a clear view of the progression of the previous VIP and future VIP-2 work, 
we use the definition of ``new'' electrons and derivation of the upper limit in the same way as we did in our past work.

In the next steps of the experiment, 
the passive shielding with circulated nitrogen gas shown schematically in Figure$~$\ref{fig:shielding} will be installed to shield the environmental gamma radiation, 
which is responsible for most of the background as discussed in section 3.3. 
Using this shielding we expect a one-order-of-magnitude reduction in background with respect to this work, 
leading to a factor of 3 improvement in the upper limit.
Since late 2017 we have been modifying the setup for a new copper strip with 25 $\mu$m thickness with shorter effective length and new SDD arrays with tripled total sensitive area, 
that will jointly increase the x-ray detection efficiency by a factor of ten, contributing directly to the improvement factor of the upper limit.  
The implementation of the modifications is planned together with the installation of the passive shielding. 
With the complete installation of the apparatus, 
the planned data taking time of 3 to 4 years will introduce a factor of 4 to 5 improvement considering the upper limit roughly scales with the inverse square-root of exposure time.
Multiplying these factors together, the VIP-2 experiment can either set a new upper limit for the probability that the PEP is violated at the level of $10^{-31}$, 
improving the VIP experiment result by two orders of magnitude, or find the PEP violation. 

\vspace{6pt}

\paragraph{Acknowledgements}
We thank H. Schneider, L. Stohwasser, and D. Pristauz-Telsnigg from Stefan-Meyer-Institut 
for their fundamental contribution in designing and building the VIP-2 setup. 
We acknowledge the very important assistance of the INFN-LNGS laboratory staff during all phases of preparation, installation and data taking.
We thank the Austrian Science Fond (FWF) which supports the VIP-2 project with the grant P25529-N20. 
We acknowledge the support from the EU COST Action CA15220, 
and from Centro Fermi (``Problemi aperti nella meccania quantistica''project).
Furthermore, this paper was made possible through the support of a grant from the John Templeton Foundation (ID 58158).
The opinions expressed in this publication are those of the authors and do not necessarily reflect the views of the John Templeton Foundation.

\begin{table*}[h]
\caption{ Energy shift and transition probability of vPp-atomic transitions for copper.}
\centering
\label{tab:calc}
\begin{tabular}[]{c | c c | c c | c}
\hline
Transition & Experimental  &  Calculated    &  \multicolumn{2}{c}{Calculated vPp: Energy (eV) and}    &    Calc. energy diff.    \\
           &  energy (eV)  &  non-vPp (eV)  &  \multicolumn{2}{c}{Transition probability (s$^{-1}$)} &     non-vPp - vPp (eV)    \\
\hline
$2p_{1/2}\rightarrow 1s_{1/2}$ ($K_{\alpha2}$)   &  8027.83  & 8027.85   &  7728.92  &   2.5690970E+14  &  298.93                \\
$2p_{3/2}\rightarrow 1s_{1/2}$ ($K_{\alpha1}$)   &  8047.78  & 8047.79   &  7746.73  &   2.6372675E+14  &  301.06                \\
$3p_{3/2}\rightarrow 1s_{1/2}$ ($K_{\beta1}$)    &  8905.29  & 8905.41   &  8531.69  &   2.6737747E+13  &  373.72                \\
$3d_{5/2}\rightarrow 2p_{3/2}$ ($L_{\alpha1}$)   &   929.70  &  929.72   &   822.83  &   3.4922759E+08  &  106.89                \\
$3d_{1/2}\rightarrow 2p_{1/2}$ ($L_{\beta1}$)    &   949.80  &  949.84   &   841.91  &   3.0154308E+08  &  107.93                \\
\hline
$3d\rightarrow 1s$ (DRR)                         &  8979.00  & 8977.14   &  8570.82  &   1.2125697E+06  &  406.32                \\
\hline
\end{tabular}
\end{table*}

\appendix

\section*{Appendix : Calculation of the PEP violating transition energy in copper}
The calculation of the Pauli-forbidden radiative-transition energy was performed through the MCDFGME numerical code \cite{Des}. 
This program solves the multiconfiguration Dirac-Fock equations self-consistently, taking into account relativistic effects. 
Specifically, the effects of the Breit operator, the Lamb shift and radiative corrections (vacuum polarization and the Uehling potential) are all included in the calculation that however excludes the formation of electron-positron pairs (no-pair approximation). 
Muons and electrons can be treated in an analogous way in the self-consistent field theory. The Dirac-Fock method employed in the MCDFGME code, as described in \cite{Mal78}, 
is a variational procedure based on a multiconfiguration approximation to the N-electron wave function. 
In the first step, a functional form for the wave function is defined in terms of a linear combination of a hydrogen-like basis. 
The coefficients of the linear combination act as variational parameters. Then an expression for the total energy is derived in terms of these parameters, and their values are obtained through energy minimization. 
Usually, a further constraint is imposed, i.e., that the total N-electron wavefunction be written as a Hartree-Fock combination of Slater determinants, 
in order to satisfy the total antisymmetrization with respect to all electron variables. However, in the present case, this condition is relaxed for one electron, that is supposed to violate the Pauli principle (vPp). 
Therefore, this vPp electron is treated by the program as a test particle of the same mass and spin as the electron, 
whose wavefunction is not antisymmetrized with those of the other electrons (as if it were a ``light'' muon). 
As a result, the exchange integrals between the vPp electron and other electronic states vanish and the vPp electron can jump to any of the atomic shells, even doubly occupied.

In more details, the total wavefunction of the N-1 ``normal'' electrons is written as an antisymmetrized sum of spin-orbital, orthogonal products of the usual form: 
$\phi_{nlm_{\sigma}}(r,\theta,\phi) = R(nl;r)Y_{lm}(\theta,\phi)X_{\sigma}$, with $Y_{lm}(\theta,\phi)$ a spherical harmonic, $X_{\sigma}$ the spin function and $R(nl;r)$ the radial function. 
The total wavefunction is an eigenfunction of the total angular momentum operators $L^2$, $L_z$ as well as the total spin operators $S^2$ and $S_z$. 
Notice that the latter constraint is actually valid at the atomic level, because of the spherical symmetry of the atom, and only approximately for copper in solid state (see below). 
The overlap between the initial and final states is taken into account and the lack of orthogonality between the initial and final states is treated by the MCDFGME code so that all orbitals are evaluated self-consistently. 
The effect of orbital relaxation in excited states is also taken into account. The program generates all the $jj$ configurations arising from a given LS configuration and builds the eigenstates of the total angular momentum, 
for which the total energy is computed. Finally, we used relativistic many-body perturbation theory for the calculations of standard (non-vPp) transitions matrix-elements. 
The magnetic-dipole transition rates between the initial and final states were found about four orders of magnitude less than the electric dipole transitions and therefore neglected.
The results of these calculations for the case of copper are reported in Table \ref{tab:calc}.

We first remark the excellent agreement of calculated and experimental energy values for ``normal'' (non-vPp) transitions (2nd and 3rd columns), 
at the level of 10$^{-6}$ in the relative error for the $K_{\alpha1}$ line, in keeping with standard calculations in this domain \cite{Ind88}. 
This shows that, compared to the energy-uncertainty scale of our detectors, the atomic calculation does not introduce any relevant source of energy-uncertainty in the estimate of the transition energy. 
We also remark that, for the $K_{\alpha2}$ transition, the transition probability for vPp electrons gives $\Delta t \sim 3.79 10^{-15}$ s, implying a natural width for the transition of $\Delta E \sim 1.1$ eV. 
Table \ref{tab:calc} provides us with the energy window for such vPp transitions: at 7729 eV for $K_{\alpha 2}$ and 7747 eV for $K_{\alpha 1}$, with an energy-shift of about 300 eV compared to their standard transition energy.

Finally, we remark that our vPp electron in the conduction $3d$ band has two possible channels to de-excite to the $1s$ state, 
either through a direct radiative transition (DRR) from the $3d$ band or through a cascade from $3d$ to $2p$ states ($L_{\alpha,\beta}$ lines) and then to $1s$, through $K_{\alpha}$ transitions. 
From the last line of Table \ref{tab:calc} we see that a DRR would take place in a time of the order of the $\mu$s, whereas the cascade is much faster, mainly determined by the 3 ns of the $L_{\alpha,\beta}$ transition,  
the $K_{\alpha}$ being practically instantaneous compared to it. This justifies the choice of VIP setup to focus on the $K_{\alpha}$ transitions, characterized by a faster transition rate of about 300 compared to the DRR.


\bibliographystyle{elsarticle-num}

\end{document}